\documentclass{article}

\PassOptionsToPackage{numbers,square}{natbib}

\usepackage[final]{neurips_2024}

\usepackage[utf8]{inputenc} %
\usepackage[T1]{fontenc}    %
\usepackage{hyperref}       %
\usepackage{url}            %
\usepackage{booktabs}       %
\usepackage{amsfonts}       %
\usepackage{nicefrac}       %
\usepackage{microtype}      %
\usepackage{xcolor}         %

\title{Position: Challenges and Opportunities for Differential Privacy in the U.S. Federal Government}

\author{%
  Amol Khanna \\ 
  Booz Allen Hamilton \\
  Boston, MA \\ 
  \texttt{Khanna\_Amol@bah.com} \\ 
  \And
  Adam McCormick \\ 
  Booz Allen Hamilton \\ 
  Washington, DC \\ 
  \texttt{McCormick\_Adam@bah.com} \\ 
  \And
  Andre Nguyen \\ 
  Booz Allen Hamilton \\ 
  Washington, DC \\ 
  \texttt{Nguyen\_Andre@bah.com} \\ 
  \And 
  Chris Aguirre \\ 
  Booz Allen Hamilton \\ 
  Austin, TX \\ 
  \texttt{Aguirre\_Chris@bah.com} \\
  \And 
  Edward Raff \\ 
  Booz Allen Hamilton \\ 
  Syracuse, NY \\ 
  \texttt{Raff\_Edward@bah.com} \\
}

\begin{document}

\maketitle

\begin{abstract}
    In this article, we seek to elucidate challenges and opportunities for differential privacy within the federal government setting, as seen by a team of differential privacy researchers, privacy lawyers, and data scientists working closely with the U.S. government. After introducing differential privacy, we highlight three significant challenges which currently restrict the use of differential privacy in the U.S. government. We then provide two examples where differential privacy can enhance the capabilities of government agencies. The first example highlights how the quantitative nature of differential privacy allows policy security officers to release multiple versions of analyses with different levels of privacy. The second example, which we believe is a novel realization, indicates that differential privacy can be used to improve staffing efficiency in classified applications. We hope that this article can serve as a nontechnical resource which can help frame future action from the differential privacy community, privacy regulators, security officers, and lawmakers. 
\end{abstract}

\section{Differential Privacy}

Despite executive orders and guidance encouraging the use of revolutionary privacy enhancing technologies to reduce the risk related to the rise of big data, artificial intelligence (AI), and machine learning (ML), institutional barriers impede the widespread deployment of differential privacy throughout the U.S. federal government. Released over the past year, “Executive Order on the Safe, Secure, and Trustworthy Development and Use of Artificial Intelligence,” and NIST SP 800-226 “Evaluating Differential Privacy Guarantees” both explore the need for adopting privacy enhancing technology to mitigate privacy risks arising from increased data processing \cite{biden2023executive,near2023guidelines}.  

Differential privacy, a privacy-preserving algorithmic framework, is one such privacy preserving technology that can be applied to a wide variety of common government agency use cases and has already been studied, documented, and deployed in commercial industry. Differential privacy measures the privacy risk to individuals by comparing the output of an algorithm with an individual’s data to the output without that data. If the output can be heavily swayed by a single datapoint, the algorithm can betray the privacy of specific individuals. If this cannot happen, the output does not indicate if a datapoint was used as an input \cite{near2021programming}. Differential privacy employs quantitative parameters to set the amount that an output can change with each additional datapoint, and by modifying these parameters, practitioners can set the allowable privacy risk of an algorithm. Differential privacy is the only existing privacy enhancing technology which can make a statistical guarantee of data privacy, indicating that released statistics, synthetic datasets, and even machine learning models cannot be reverse engineered to identify individuals in their source datasets. Unlike other methods, the privacy guarantee is not conditional and cannot be attacked, which makes it an attractive option for guaranteeing privacy in privacy-sensitivity applications \cite{dwork2006calibrating}.  

\section{Attempts and Challenges of Federal Differential Privacy Implementations}

Within the federal government, differential privacy has the potential to reduce potential harms caused by data leaks and enable more public releases of data and statistics, which can lead to better informed political and economic decisions \cite{krishnamurthy2016liberating}. Indeed, the U.S. Census Bureau has already taken steps in this direction by using differential privacy to mask some of the sensitive statistics it released from the 2020 Census \cite{abowd2018us}. By using differential privacy, the Bureau guaranteed that specific households cannot be identified from the privatized statistics while still releasing information to lawmakers \cite{su20242020unitedstatesdecennial}. 

Despite the U.S. government touting the benefits of differential privacy and encouraging its adoption for a variety of use cases, the use of the technology has largely been restricted, to wit there are no other prominent examples. We argue that the limited government usage of the technology can be attributed to a lack of awareness among government program managers, challenging deployments in large-scale systems, and most critically, unclear guidance to security officers challenging deployments in large-scale systems. We discuss each of these challenges in turn: 
\begin{enumerate}
    \item Lack of Awareness: differential privacy is a statistical method for guaranteeing privacy, which departs from older privacy-preservation techniques that were deterministic, but ineffective. Older methods, like record anonymization, removing sensitive attributes, and K-anonymity, operated directly on datasets \cite{sweeney2002k}. These methods could typically be used as a dataset preprocessing step prior to using a standard algorithm for statistics or machine learning. However, these methods are still very vulnerable to privacy attacks; indeed, government datasets that employed the aforementioned methods have been attacked in the past \cite{barth2012re}. Differential privacy is robust to all attacks, but effectively using it requires a deep understanding of probability and statistics, which many privacy professionals are unfamiliar with. Additionally, differential privacy’s statistical nature often requires modifying an entire algorithmic pipeline; instead of simply preprocessing the rows of a dataset, significant modifications must be made to a target algorithm to produce useful results \cite{ngong2024evaluating}. This statistical nature can also impact decision-makers, as the notion that data is secure but probabilistically protected is a departure from better-understood methods like cryptography. Finally, current implementations of differential privacy use functional programming APIs, which can be less accessible to data scientists than more common object-oriented frameworks \cite{zhang2023evaluation}. 
    \item Challenging Deployments: differential privacy can be challenging to use for tasks with multiple goals and intermediate computations \cite{zhao2022survey}. In such tasks, there are many potential statistical missteps that can produce sources of privacy leaks. Large technology companies have approached this challenge by focusing on deploying differential privacy for targeted tasks and including differential privacy experts to oversee deployments. However, government agencies are often tasked with releasing social and political information to the public, and these datasets often require multiple sources and released values. 
    \item Unclear Guidance: most importantly, many overseeing security officers in the government are reluctant to use differential privacy since no government authority has released official approvals for using the technology in privacy-critical tasks. Since bureaucratic approvals exist for older technologies, security officers choose to use these technologies despite their increased susceptibility to privacy attacks such as membership inference and model inversion attacks \cite{OTCNet, Hashing}. We believe that for differential privacy to permeate through government applications, a government authority must approve its use in multiple privacy-critical applications, since this can set a government standard. 
\end{enumerate} 

Each of these challenges was realized in the Census Bureau's differential privacy deployment. First, approvals for differential privacy use did not exist, and the Census used privacy-enhancing dataset preprocessing techniques in conjunction with differential privacy \cite{garfinkel2022differential}. Next, a number of released values had to be modified since the Census contains many codependent statistics, and adding noise to many individual statistics can significant diminsh the utility of the final result. Finally, many statisticians and social scientists critiqued the Bureau's use of differential privacy for having multiple potential privacy leaks and discrepancies among statistics used to inform political and economic decisions \cite{mueller20222020,boyd2022differential,hawes2020implementing,cohen2022private}. This was exacerbated by the fact that differentially private noise disproportionately affects underrepresented communities in datasets, and social scientists analyzing Census data often seek to identify and help marginalized communities. Ultimately, the Census Bureau has made a massive step forward in the practice of real-world differential privacy, and securing citizens' information, which are hard-fought lessons we wish to see replicated in other agencies. 

\section{Differentiating Private Releases by Perceived Trustworthiness}

As a state-of-the-art privacy enhancing technology, we believe that differential privacy should be incorporated into more government agency applications. In addition, we argue that differential privacy’s parametric approach enables data security officers to further improve customizability and efficiency in government applications. Specifically, security officers can inject their knowledge of a dataset’s sensitivity and an algorithm’s end users through differential privacy’s quantitative parameters \cite{nanayakkara2023chances}. This customization means that different versions of the same algorithm can be released with varying levels of privacy based on a perceived level of risk. 

For example, if an agency working with personal health information built a machine learning model on patient records, the agency could enable physicians to access a less private and more accurate model while releasing a more private version to academics and insurance companies. This is because the agency may see registered physicians as a trusted group, while researchers and insurance companies are unknown public entities. By releasing the two different versions of the model, the agency enables physicians to access a state-of-the-art model for disease diagnosis and treatment while spurring advancements in research and policy with a more private model. Using other privacy enhancing methods without this fine-grained control may have resulted in a less accurate model for physicians or a health data breach, either of which are unsatisfactory outcomes. 

\section{Improving Efficiency with Privacy: An Unrealized Frontier}

The previous section's example highlights differential privacy’s applicability in civilian agencies. We also argue that differential privacy could enable significant efficiency gains in classified settings, and we believe that we are the first to identify this opportunity. Government defense agencies often generate statistics and models on datasets classified, e.g., as Top Secret. Under current guidance, every person with access to any part of these modeling pipelines, such as data annotators, data scientists, machine learning engineers, software engineers, deployment specialists, and even eventual users, must have a Top Secret clearance. This is because all of these individuals can either access raw Top Secret data or derivatives of this data, which could be reverse-engineered into the Top Secret data. However, staffing technical roles at such a high clearance level is challenging due to a shortage of cleared staff, and as such, agencies often choose not to pursue projects in highly classified settings due to a lack of cleared and capable personnel \cite{groeber2020federal,berger2019us,farrell2017personnel}. We believe that if differential privacy was used when creating data derivatives, certain positions in these pipelines could be staffed at lower clearance levels. For example, if a sufficient level of privacy was used when generating statistical reports on Top Secret datasets, software engineers and deployment specialists would be unable reverse engineer these reports to identify specific Top Secret datapoints. Thus, there is a reduced risk associated with these reports, meaning that personnel handling, deploying, and using these reports can have lower clearance levels. Staffing at lower clearance levels is significantly easier and cheaper, so adopting this policy on highly classified projects can significantly reduce time and monetary costs and allow agencies to pursue more initiatives. 

\section{Conclusion}

This work seeks to communicate the challenges our team of differential privacy researchers, privacy lawyers, and data scientists have had when communicating about differential privacy to decision-makers at government agencies. We have noticed that senior data scientists are wary of using the technology due to its significant departure from previous methods and its difficulty in scaling to large applications and unstructured datasets. Security officers often avoid novel privacy preserving technologies due to a lack of government guidance and specifications on deployments. We note that each of these challenges has been discussed in prior literature on differential privacy, but is particularly relevant to the highly structured and regulated setting of data processing at government agencies \cite{franzen2022private,garfinkel2018issues}. 

We then proceed to highlight two examples of government tasks that could be improved with differential privacy. The first demonstrates that in civilian agencies, the government can use the quantitative nature of differential privacy to release different versions of the same output to different populations based on a perceived level of privacy risk. This can enable government agencies to serve each stakeholder of their released statistics and models effectively while managing the privacy risk of sensitive data. Our next use case highlights that in classified settings, using differential privacy can allow the government to staff positions handling data derivatives at lower clearance levels, which can significantly improve efficiency. We note that we have not seen this use case discussed in any other literature; this is likely because classification and clearances are only strongly enforced in  government settings. As such, we hope that by communicating this use case clearly, government security officers can consider lowering classifications of data derivatives when using differential privacy. 

Finally, we conclude by highlighting four areas of future action critical to the success of deploying differential privacy in government applications. 
\begin{enumerate}
    \item The differential privacy community must develop an effective method for communicating privacy guarantees to security officers and decision makers. The community is already aware of this critical need and recent research works on this problem \cite{ngong2024evaluating,cummings2023centering}. 
    \item Differential privacy researchers should develop better methods for privatizing large, unstructured datasets. These forms of datasets are underexplored in the differential privacy literature but are very common in government applications. 
    \item There is a significant gap between theoretical works on differential privacy and practical differential privacy deployments. We identify works authored in-part by government employees focused on empirical methods for differential privacy, and believe that government agencies would be best served by empirical works and implementations of theoretical methods as these provide evidence that differential privacy is ready for deployment \cite{sen2024diverse,khanna2023differentially,khanna2024sok,leblond2023probing,lu2022general,Swope24,dpScreeningHard}.  
    \item Regulators, lawmakers, and security officers must decide on a framework for evaluating, accepting, and deploying differentially private systems on government datasets. 
\end{enumerate}
We believe that effectively addressing these three challenges will unlock significant use cases for differential privacy in the government setting.

\bibliographystyle{unsrtnat}
\bibliography{main}

\begin{thebibliography}{33}
\providecommand{\natexlab}[1]{#1}
\providecommand{\url}[1]{\texttt{#1}}
\expandafter\ifx\csname urlstyle\endcsname\relax
  \providecommand{\doi}[1]{doi: #1}\else
  \providecommand{\doi}{doi: \begingroup \urlstyle{rm}\Url}\fi

\bibitem[Biden(2023)]{biden2023executive}
Joseph~R Biden.
\newblock Executive order on the safe, secure, and trustworthy development and use of artificial intelligence.
\newblock 2023.

\bibitem[Near et~al.(2023)Near, Darais, Lefkovitz, Howarth, et~al.]{near2023guidelines}
Joseph~P Near, David Darais, Naomi Lefkovitz, Gary Howarth, et~al.
\newblock Guidelines for evaluating differential privacy guarantees.
\newblock \emph{National Institute of Standards and Technology, Tech. Rep}, 2023.

\bibitem[Near and Abuah(2021)]{near2021programming}
Joseph~P Near and Chik{\'e} Abuah.
\newblock Programming differential privacy.
\newblock 2021.

\bibitem[Dwork et~al.(2006)Dwork, McSherry, Nissim, and Smith]{dwork2006calibrating}
Cynthia Dwork, Frank McSherry, Kobbi Nissim, and Adam Smith.
\newblock Calibrating noise to sensitivity in private data analysis.
\newblock In \emph{Theory of Cryptography: Third Theory of Cryptography Conference, TCC 2006, New York, NY, USA, March 4-7, 2006. Proceedings 3}, pages 265--284. Springer, 2006.

\bibitem[Krishnamurthy and Awazu(2016)]{krishnamurthy2016liberating}
Rashmi Krishnamurthy and Yukika Awazu.
\newblock Liberating data for public value: The case of data. gov.
\newblock \emph{International Journal of Information Management}, 36\penalty0 (4):\penalty0 668--672, 2016.

\bibitem[Abowd(2018)]{abowd2018us}
John~M Abowd.
\newblock The us census bureau adopts differential privacy.
\newblock In \emph{Proceedings of the 24th ACM SIGKDD international conference on knowledge discovery \& data mining}, pages 2867--2867, 2018.

\bibitem[Su et~al.(2024)Su, Su, and Wang]{su20242020unitedstatesdecennial}
Buxin Su, Weijie~J. Su, and Chendi Wang.
\newblock The 2020 united states decennial census is more private than you (might) think, 2024.

\bibitem[Sweeney(2002)]{sweeney2002k}
Latanya Sweeney.
\newblock k-anonymity: A model for protecting privacy.
\newblock \emph{International journal of uncertainty, fuzziness and knowledge-based systems}, 10\penalty0 (05):\penalty0 557--570, 2002.

\bibitem[Barth-Jones(2012)]{barth2012re}
Daniel Barth-Jones.
\newblock The're-identification'of governor william weld's medical information: a critical re-examination of health data identification risks and privacy protections, then and now.
\newblock \emph{Then and Now (July 2012)}, 2012.

\bibitem[Ngong et~al.(2024)Ngong, Stenger, Near, and Feng]{ngong2024evaluating}
Ivoline~C Ngong, Brad Stenger, Joseph~P Near, and Yuanyuan Feng.
\newblock Evaluating the usability of differential privacy tools with data practitioners.
\newblock In \emph{Twentieth Symposium on Usable Privacy and Security (SOUPS 2024)}, pages 21--40, 2024.

\bibitem[Zhang et~al.(2023)Zhang, Hagermalm, Slavnic, Schiller, and Almgren]{zhang2023evaluation}
Shiliang Zhang, Anton Hagermalm, Sanjin Slavnic, Elad~Michael Schiller, and Magnus Almgren.
\newblock Evaluation of open-source tools for differential privacy.
\newblock \emph{Sensors}, 23\penalty0 (14):\penalty0 6509, 2023.

\bibitem[Zhao and Chen(2022)]{zhao2022survey}
Ying Zhao and Jinjun Chen.
\newblock A survey on differential privacy for unstructured data content.
\newblock \emph{ACM Computing Surveys (CSUR)}, 54\penalty0 (10s):\penalty0 1--28, 2022.

\bibitem[Team(2020)]{OTCNet}
OTCnet~Deployment Team.
\newblock Otcnet pii information, 2020.

\bibitem[in~the Office~of Technology(2024)]{Hashing}
Staff in~the Office~of Technology.
\newblock No, hashing still doesn't make your data anonymous, 2024.

\bibitem[Garfinkel(2022)]{garfinkel2022differential}
Simson Garfinkel.
\newblock Differential privacy and the 2020 us census.
\newblock 2022.

\bibitem[Mueller and Santos-Lozada(2022)]{mueller20222020}
J~Tom Mueller and Alexis~R Santos-Lozada.
\newblock The 2020 us census differential privacy method introduces disproportionate discrepancies for rural and non-white populations.
\newblock \emph{Population Research and Policy Review}, 41\penalty0 (4):\penalty0 1417--1430, 2022.

\bibitem[Boyd and Sarathy(2022)]{boyd2022differential}
Danah Boyd and Jayshree Sarathy.
\newblock Differential perspectives: Epistemic disconnects surrounding the us census bureau’s use of differential privacy.
\newblock \emph{Harvard Data Science Review (Forthcoming)}, 2022.

\bibitem[Hawes(2020)]{hawes2020implementing}
Michael~B Hawes.
\newblock Implementing differential privacy: Seven lessons from the 2020 united states census.
\newblock \emph{Harvard Data Science Review}, 2\penalty0 (2):\penalty0 4, 2020.

\bibitem[Cohen et~al.(2022)Cohen, Duchin, Matthews, and Suwal]{cohen2022private}
Aloni Cohen, Moon Duchin, J~Matthews, and Bhushan Suwal.
\newblock Private numbers in public policy: Census, differential privacy, and redistricting.
\newblock \emph{Harvard Data Science Review}, \penalty0 (Special Issue 2), 2022.

\bibitem[Nanayakkara et~al.(2023)Nanayakkara, Smart, Cummings, Kaptchuk, and Redmiles]{nanayakkara2023chances}
Priyanka Nanayakkara, Mary~Anne Smart, Rachel Cummings, Gabriel Kaptchuk, and Elissa~M Redmiles.
\newblock What are the chances? explaining the epsilon parameter in differential privacy.
\newblock In \emph{32nd USENIX Security Symposium (USENIX Security 23)}, pages 1613--1630, 2023.

\bibitem[Groeber et~al.(2020)Groeber, Mayberry, Crosby, Doboga, Dinicola, Lee, and Tunstall]{groeber2020federal}
Ginger Groeber, Paul~W Mayberry, Brandon Crosby, Mark Doboga, Samantha~E Dinicola, Caitlin Lee, and Ellen~E Tunstall.
\newblock Federal civilian workforc e hiring, recruitment, and related compensation practices for the twenty-first century.
\newblock 2020.

\bibitem[Berger(2019)]{berger2019us}
Benjamin~F Berger.
\newblock \emph{US security clearances: Reducing the security clearance backlog while preserving information security}.
\newblock PhD thesis, Monterey, CA; Naval Postgraduate School, 2019.

\bibitem[Farrell(2017)]{farrell2017personnel}
Brenda~S Farrell.
\newblock Personnel security clearances: Additional actions needed to ensure quality, address timeliness, and reduce investigation backlog.
\newblock 2017.

\bibitem[Franzen et~al.(2022)Franzen, von Voigt, S{\"o}rries, Tschorsch, and M{\"u}ller-Birn]{franzen2022private}
Daniel Franzen, Saskia~Nu{\~n}ez von Voigt, Peter S{\"o}rries, Florian Tschorsch, and Claudia M{\"u}ller-Birn.
\newblock " am i private and if so, how many?"--using risk communication formats for making differential privacy understandable.
\newblock \emph{arXiv preprint arXiv:2204.04061}, 2022.

\bibitem[Garfinkel et~al.(2018)Garfinkel, Abowd, and Powazek]{garfinkel2018issues}
Simson~L Garfinkel, John~M Abowd, and Sarah Powazek.
\newblock Issues encountered deploying differential privacy.
\newblock In \emph{Proceedings of the 2018 Workshop on Privacy in the Electronic Society}, pages 133--137, 2018.

\bibitem[Cummings and Sarathy(2023)]{cummings2023centering}
Rachel Cummings and Jayshree Sarathy.
\newblock Centering policy and practice: Research gaps around usable differential privacy.
\newblock In \emph{2023 5th IEEE International Conference on Trust, Privacy and Security in Intelligent Systems and Applications (TPS-ISA)}, pages 122--135. IEEE, 2023.

\bibitem[Sen et~al.(2024)Sen, Task, Kapur, Howarth, and Bhagat]{sen2024diverse}
Aniruddha Sen, Christine Task, Dhruv Kapur, Gary Howarth, and Karan Bhagat.
\newblock Diverse community data for benchmarking data privacy algorithms.
\newblock \emph{Advances in Neural Information Processing Systems}, 36, 2024.

\bibitem[Khanna et~al.(2023{\natexlab{a}})Khanna, Lu, Raff, and Testa]{khanna2023differentially}
Amol Khanna, Fred Lu, Edward Raff, and Brian Testa.
\newblock Differentially private logistic regression with sparse solutions.
\newblock In \emph{Proceedings of the 16th ACM Workshop on Artificial Intelligence and Security}, pages 1--9, 2023{\natexlab{a}}.

\bibitem[Khanna et~al.(2024)Khanna, Raff, and Inkawhich]{khanna2024sok}
Amol Khanna, Edward Raff, and Nathan Inkawhich.
\newblock Sok: A review of differentially private linear models for high-dimensional data.
\newblock In \emph{2024 IEEE Conference on Secure and Trustworthy Machine Learning (SaTML)}, pages 57--77. IEEE, 2024.

\bibitem[LeBlond et~al.(2023)LeBlond, Munoz, Lu, Fuchs, Zaresky-Williams, Raff, and Testa]{leblond2023probing}
Tyler LeBlond, Joseph Munoz, Fred Lu, Maya Fuchs, Elliot Zaresky-Williams, Edward Raff, and Brian Testa.
\newblock Probing the transition to dataset-level privacy in ml models using an output-specific and data-resolved privacy profile.
\newblock In \emph{Proceedings of the 16th ACM Workshop on Artificial Intelligence and Security}, pages 23--33, 2023.

\bibitem[Lu et~al.(2022)Lu, Munoz, Fuchs, LeBlond, Zaresky-Williams, Raff, Ferraro, and Testa]{lu2022general}
Fred Lu, Joseph Munoz, Maya Fuchs, Tyler LeBlond, Elliott Zaresky-Williams, Edward Raff, Francis Ferraro, and Brian Testa.
\newblock A general framework for auditing differentially private machine learning.
\newblock \emph{Advances in Neural Information Processing Systems}, 35:\penalty0 4165--4176, 2022.

\bibitem[Swope et~al.(2024)Swope, Khanna, Doldo, Roy, and Raff]{Swope24}
Ryan Swope, Amol Khanna, Philip Doldo, Saptarshi Roy, and Edward Raff.
\newblock {Feature Selection from Differentially Private Correlations}.
\newblock In \emph{Proceedings of the 17th ACM Workshop on Artificial Intelligence and Security (AISec'24)}, 2024.
\newblock URL \url{https://arxiv.org/abs/2408.10862}.

\bibitem[Khanna et~al.(2023{\natexlab{b}})Khanna, Lu, and Raff]{dpScreeningHard}
Amol Khanna, Fred Lu, and Edward Raff.
\newblock The challenge of differentially private screening rules.
\newblock \emph{2nd AdvML Frontiers Workshop at 40th International Conference on Machine Learning}, 2023{\natexlab{b}}.
\newblock URL \url{https://arxiv.org/abs/2303.10303}.

\end{thebibliography}

\newpage

\end{document}